\documentclass[preprint2]{aastex61}

\begin{document}

\title{Extinction ratios in the inner Galaxy as revealed by the VVV survey}

\correspondingauthor{Javier Alonso-Garc\'{i}a}
\email{javier.alonso@uantof.cl}

\author[0000-0003-3496-3772]{Javier Alonso-Garc\'{i}a}
\affil{Unidad de Astronom\'{i}a, Facultad Cs. B\'{a}sicas, Universidad de Antofagasta, Avda. U. de Antofagasta 02800, Antofagasta, Chile} 
\affil{Instituto Milenio de Astrof\'{i}sica, Santiago, Chile}

\author[0000-0002-7064-099X]{Dante Minniti}
\affil{Departamento de F\'{i}sica, Facultad de Ciencias Exactas, Universidad Andr\'{e}s Bello, Av. Fernandez Concha 700, Las Condes, Santiago, Chile}
\affil{Instituto Milenio de Astrof\'{i}sica, Santiago, Chile}
\affil{Vatican Observatory, V-00120 Vatican City State, Italy} 

\author[0000-0001-6003-8877]{M\'{a}rcio Catelan}
\affil{Instituto de Astrof\'{i}sica, Pontificia Universidad Cat\'{o}lica de Chile, Av. Vicu\~{n}a Mackenna 4860, Santiago, Chile}
\affil{Instituto Milenio de Astrof\'{i}sica, Santiago, Chile}

\author{Rodrigo Contreras Ramos}
\affil{Instituto Milenio de Astrof\'{i}sica, Santiago, Chile}
\affil{Instituto de Astrof\'{i}sica, Pontificia Universidad Cat\'{o}lica de Chile, Av. Vicu\~{n}a Mackenna 4860, Santiago, Chile}

\author{Oscar A. Gonzalez}
\affil{UK Astronomy Technology Centre, Royal Observatory, Blackford Hill, Edinburgh EH9 3HJ, UK}

\author{Maren Hempel}
\affil{Instituto de Astrof\'{i}sica, Pontificia Universidad Cat\'{o}lica de Chile, Av. Vicu\~{n}a Mackenna 4860, Santiago, Chile}

\author{Philip W. Lucas}
\affil{Department of Astronomy, University of Hertfordshire, Hertfordshire, UK}

\author[0000-0001-6878-8648]{Roberto K. Saito}
\affil{Departamento de F\'{i}sica, Universidade Federal de Santa Catarina, Trindade 88040-900, Florian\'{o}polis, SC, Brazil} 

\author[0000-0002-6092-7145]{Elena Valenti}
\affil{European Southern Observatory, Karl-Schwarszchild-Str. 2, D-85748 Garching bei Muenchen, Germany} 

\author{Manuela Zoccali}
\affil{Instituto de Astrof\'{i}sica, Pontificia Universidad Cat\'{o}lica de Chile, Av. Vicu\~{n}a Mackenna 4860, Santiago, Chile}
\affil{Instituto Milenio de Astrof\'{i}sica, Santiago, Chile}

\begin{abstract}
  Interstellar extinction towards the Galactic Center is large and significantly differential. Its reddening and dimming effects in red clump stars in the Galactic Bulge can be exploited to better constrain the extinction law towards the innermost Galaxy.  By virtue of a deep and complete catalog of more than 30 million objets at $|l|\le2\fdg7$ and $|b|\le1\fdg55$ obtained from VVV survey observations, we apply the red clump method to infer the selective-to-total extinction ratios in the $Z$, $Y$, $J$, $H$ and $K_s$ broadband near-infrared filters. The measured values are smaller than previously reported, and are not constant, with mean values, e.g., $A_{K_S}/E(J-K_s)=0.428\pm0.005\pm0.04$ and $A_{K_S}/E(H-K_s)=1.104\pm0.022\pm0.2$. We also obtain a ratio $A_Z$:$A_Y$:$A_J$:$A_H$:$A_{K_S}$ of 7.74:5.38:3.30:1.88:1.0, implying extinction towards the Galactic Center to follow a distribution as a function of wavelength steeper than previously reported, consistent with a power law $A_{\lambda}\propto{\lambda}^{-2.47}$ in the near-infrared.
\end{abstract}


\section{Introduction}
\label{intro}
The dust and gas in our Galaxy can produce significant variations in the magnitudes and colors of galactic and extragalactic objects we observe, leading to inaccurate measurements of important physical parameters, such as their distances or ages. Therefore, it is of the utmost importance to know how extinction changes in our Galaxy at the different wavelengths and lines of sight. One of the most useful methods to study the extinction and its variations in the Milky Way, especially when looking towards the inner parts of the our Galaxy where both reddening and stellar density are higher, is the so-called red clump (RC) method \citep{nis06a}.

In the color-magnitude diagrams (CMDs) of the bulge of our Galaxy, the RC is a prominent feature. RC stars are core-helium burning stars, that also burn hydrogen in a shell. They differ from horizontal branch (HB) stars by having usually higher metallicities and a buffer of mass above the H-burning shell that allows them to burn He at lower effective temperatures \citep{gir16}. Their ubiquity and their well-known intrinsic properties, which are well calibrated by models, have promoted their use as standard candles. Potential problems in this use are significantly reduced when studying the bulge of our Galaxy at near-infrared wavelengths \citep{gir16}: the use of the $K_s$ filter makes the RC highly resilient to variations in metallicity; the study of the Galactic Bulge, which engulfs a mostly old population, results in most of the stars in the RC to be low-mass stars that have undergone a Helium flash where they end up with the same core mass, which produces a more concentrated clump where differences in the envelope masses and compositions should only generate small differences in color and brightness. Therefore, the RC method establishes that for the case of RC stars located at the same distance, their location in the CMD should follow a straight line with slope $A_{\lambda}/E(M_{\lambda'}-M_{\lambda})$ in accordance with the variable extinction to each of them \citep{nis06a}.  

The VISTA Variables in the V\'ia L\'actea (VVV) survey \citep{min10} and its extension, the VVV eXtendend (VVVX) survey, have been observing the inner regions of our Galaxy in the near-infrared since 2010. The VVV survey has provided the most complete atlas of the stellar populations in the inner Milky Way, and most of the bulge stars in this atlas are RC stars \citep{sai12a}. \citet{sai11} and \citet{gon11a,gon15} have used the RC stars found in VVV and in 2MASS to trace the structure of the Galactic Bulge and bar. \citet{gon11b,gon12} have used them to trace the extinction providing a 2D high-resolution color excess map of the Galactic Bulge. \citet{che13} have extended this study to provide a 3D map.

The reddening law has been found to be non-standard towards the low-latitude inner regions of our Galaxy \citep{nat13}, where it is better described by smaller total-to-selective ratios than the canonical values provided in \citet{car89} or \citet{rie85}, as measured in the optical \citep{dra03,uda03} and in the infrared \citep{nis06a,nis09}. Specifically in the near- and mid-infrared, the RC method has been extensively used in the inner area of the Galaxy to study the behavior of the extinction ratios, but the very center square degrees of the Galaxy were omitted from the analysis, since observations were not deep enough to sample the RC there. For the same reason, the color baseline was not complete and the RC method could not be employed to its full potential.

In this letter we make use of a new catalog of VVV sources obtained from PSF photometry analysis to find the selective-to-total extinction ratios $A_{\lambda}/E(M_{\lambda'}-M_{\lambda})$ towards the inner Galaxy, for the different combinations of near-infrared filters available in the VVV survey. This new, deeper, and more complete catalog allows us to observe highly reddened RC stars not available to previous studies, which lets us include the most central, highly reddened and crowded areas in the Milky Way in our analysis.

\section {Observations and data reduction}
VVV observations are taken with the 4.1m VISTA telescope in the Cerro Paranal Observatory in Chile, in five near-infrared filters ($Z$, $Y$, $J$, $H$, and $K_S$). The VISTA camera, VIRCAM, contains 16 chips with significant gaps between them, which provide a non-contiguous coverage of $1.5\times1.1$ square degrees in the sky, and produces so-called pawprint images. VVV observing strategy, detailed in \citet{sai12a}, is to take firstly a set of 2 consecutive, slightly-offset ($\sim20 \arcsec$) images that, when combined in a so-called stacked pawprint images, allow to clean some of the cosmetic defects of the chips, in addition to provide a deeper observation. A mosaic of six of these stacked pawprint images, observed in a pattern to cover the gap regions among chips, is later obtained and combined in so-called tile images, to give a complete coverage of everyone of the 348 VVV fields.

\begin{figure*}
\plottwo{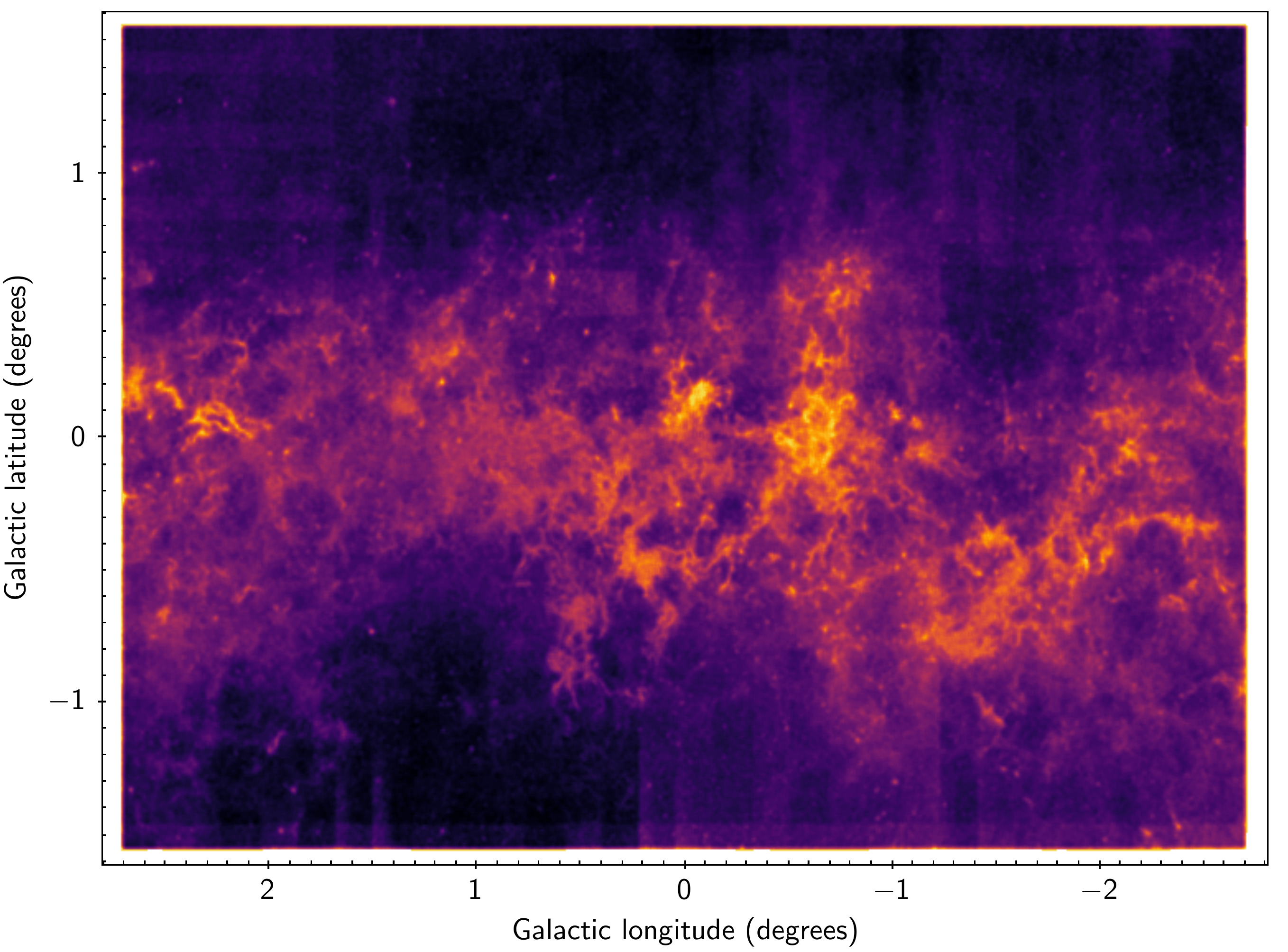}{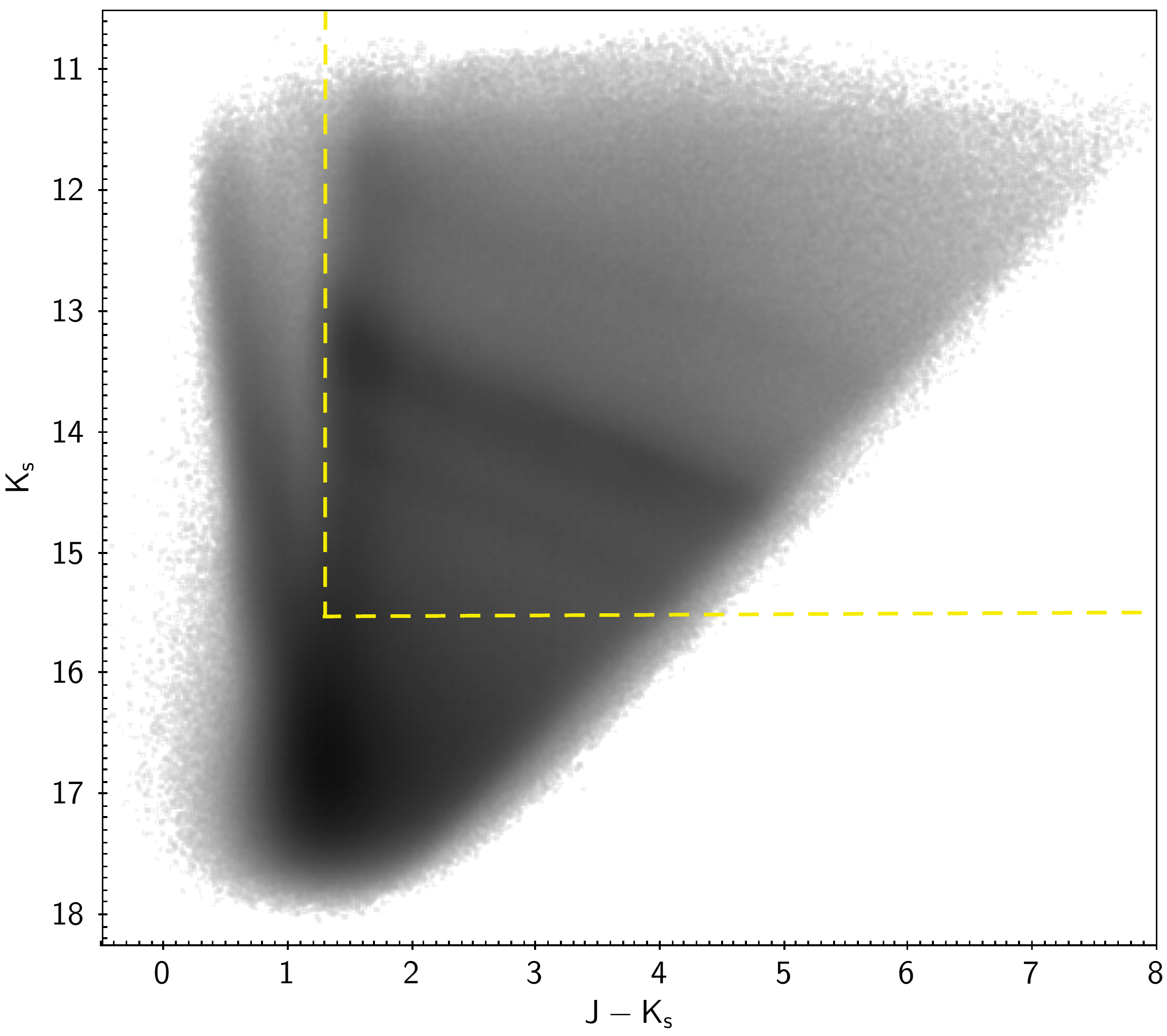}
\caption{On the left, density map of the studied area towards the GC. Colder colors mean higher stellar densities in those regions. The millions of stars detected with our PSF photometry pipeline allow for a high-definition map of the reddening. On the right, our $(J-K_s)$ vs. $K_s$ CMD of the same area. Main sequence disk stars dominate at colors bluer than $(J-K_s)=1.3$, while Galactic Bulge giants are dominant at colors redder than $(J-K_s)=1.3$, and magnitudes brighter than $K_s\sim15.5$. Differential reddening effects are the main factor contributing to the broadening and dimming of the different evolutionary sequences, especially the bulge giants.}
\label{fig_all}
\end{figure*}

For our analysis, we use VVV observations of the galactic central regions with $|l|\le2\fdg7$ and $|b|\le1\fdg55$. They encompass 12 VVV fields (b318 to b321, b332 to b335, and b346 to b349). We extract the photometry of the stars in the 12 target VVV fields in the $Z$, $Y$, $J$, $H$, and $K_S$ near-infrared filters. Details on the photometry extraction on these and the rest of the VVV fields will be presented in a future paper. Suffice it to say here that we identify sources and extract their PSF photometry, using the DoPHOT software package \citep{sch93,alo12}, on every one of the individual chips of the camera on the VISTA telescope available from the stacked pawprint images \citep{sai12a,alo15} for the 12 fields of interest, in 2 epochs, for the 5 near-infrared filters available. Working with individual chips on the stacked pawprints allows us to avoid complications in the modeling of the PSF present in other data-products of the VVV survey \citep{alo15}. Photometry is astrometrized and calibrated into the VISTA photometric system by comparison with the dimmer aperture photometry provided by CASU \citep{eme04,ham04,irw04}. Photometry in every filter of the individual chips and fields is later cross-correlated according to the positions of the sources in the sky using STILTS \citep{tay06}. For every filter, we keep the photometry only if the object appears in the two epochs per field analyzed. We cross-correlated again the photometry in all the filters available, and retain it only for objects which appear in at least 3 of the 5 near-infrared filters available. This way we get rid of most spurious detections. Finally, the VVV field disposition follows a distribution according to galactic longitude and latitude, but it is not completely symmetrical with respect to them \citep{min10,sai12a}. For this work, we prefer to perform an analysis with a symmetric coverage in galactic coordinates in the area defined in the beginning of this section, so we do not use all the area covered by the previously-mentioned target VVV fields at most positive latitudes and longitudes. The resulting catalog contains more than 31 million objects in the central galactic region studied. It reaches deeper and has a higher level of completeness than the VVV photometric catalogs currently available publicly, which allow us to almost triple the number of detected RC stars. The stars identified in our new catalog are irregularly distributed as it is shown in the left panel in Figure~\ref{fig_all}. These variations in the densities of stars are produced by the significant presence of gas and dust in our line of sight, which produces rapid changes in the completeness and detection limit for almost contiguous sections of the sky, even in the near-infrared wavelengths where VVV observations are taken.

\section{Analysis}
We build the CMDs for the different combinations of filters available. In the right panel in Figure~\ref{fig_all}, we can clearly observe the combination of main-sequence disk stars and giant bulge stars present, as described in \citet{sai12a}, although evolutionary sequences are broadened and dimmed considerably, mainly for the bulge stars. As variations in distance due to the X-shape of the Galactic Bulge should not be significant at these small galactic latitudes and longitudes \citep{gon15}, the main responsible cause for this effect should be extinction. If we assume the same distribution in metallicities and distances for the RC stars in the relatively limited sky region included in our study, we can analyze the behavior of the extinction just by locating the positions of this feature at different positions in the CMD.

\begin{figure*}
\plotone{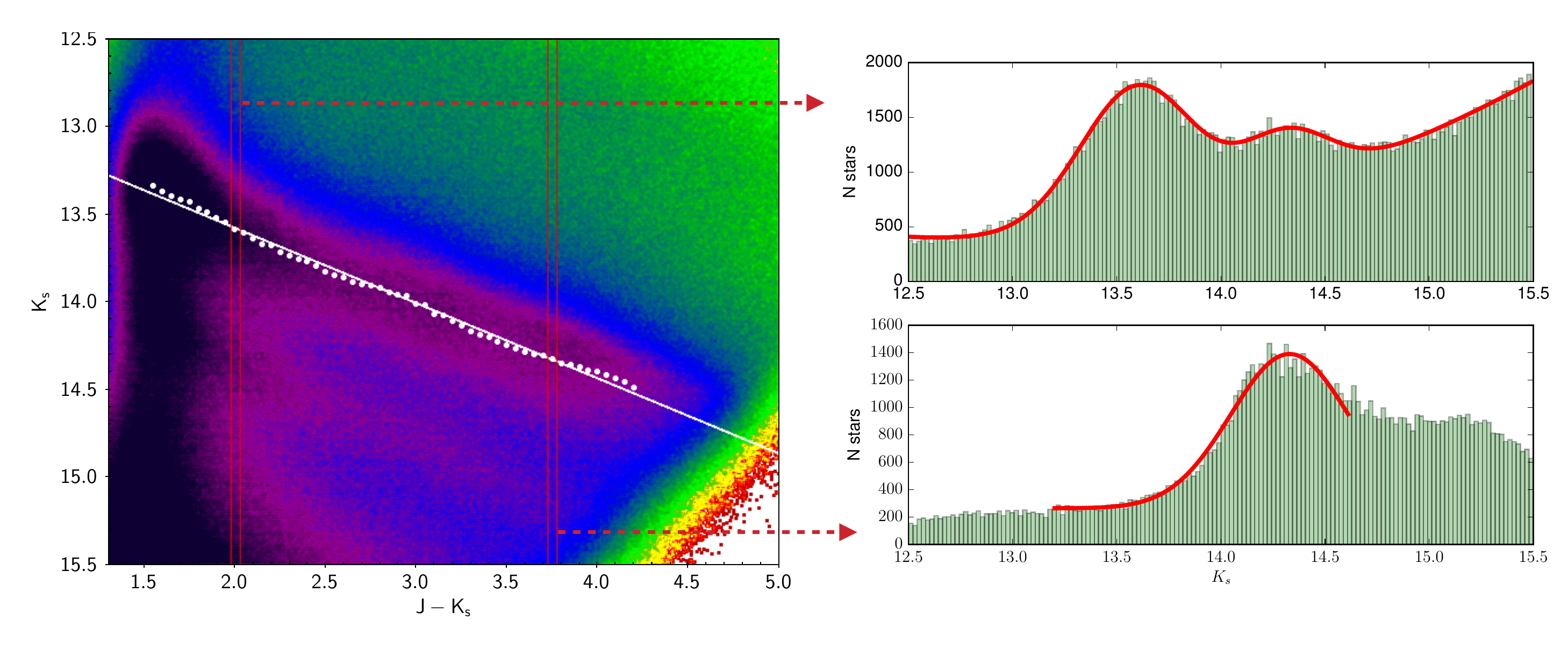} 
\caption{On the left, section of the $(J-K_s)$ vs. $K_s$ CMD where most of the giant stars from the Galactic Bulge are located. Colder colors mean higher densities of stars. In our analysis, the CMD is divided in small sections according to color. Red boxes show two of these subsections as examples. Histograms are produced for every subsection, as shown in the right panels. The red lines in the histograms show the fit we perform to every histogram distribution, as described in the text. This way we are able to identify the RC and secondary bump positions (upper right panel) or RC for sections with lower completeness (lower right panel). The RC positions identified this way are plotted in the left panel CMD as white dots, and then a linear fit to them allows to calculate the selective-to-total extinction ratio.}
\label{fig_cmd}
\end{figure*}

Differently from previous similar analyses \citep{nis06a,nis09}, we do not divide the analyzed sky area in smaller cells to look for the RC position. Instead, we divide our CMDs in narrow, 0.05 magnitudes wide, color sections (see left panel in Figure~\ref{fig_cmd}), generate histograms of the stars present in these color cells as shown in the right panels of Figure~\ref{fig_cmd}, and try to fit them with a second order polynomial function plus two gaussians, which have been shown to accurately represent the distribution of RGB, RC, and secondary bump stars\footnote{The secondary bump has been identified with the bulge red giant branch bump \citep{weg13,nat13}, although it may correspond to another different feature \citep{gon11a}.}, respectively \citep{gon11b,gon13,weg13}. The fit is generally good (upper right panel in Figure~\ref{fig_cmd}), although as we move towards redder colors, incompleteness at the dimmest magnitudes start to increase until it prevents a proper fit. When this happens, and since the RC overdensity is still significant in the histogram for redder colors, we decided to just fit a gaussian plus some constant in the histogram of our brighter stars at a given color (lower right panel in Figure~\ref{fig_cmd}), to find the bulge RC position at the reddest colors available (see left panel in Figure~\ref{fig_cmd}). As shown in \citet{bab05}, $\sigma_{RC}$, which measures the dispersion of the gaussian fit, is the convolution of the true line-of-sight dispersion in distance and metallicity of the RC stars, the intrinsic dispersion of the RC luminosity, and photometric errors. The values of $\sigma_{RC}$ in our fits in $K_s$ are between 0.25 and 0.3 mag, which agrees with the values measured in other studies for regions at low Galactic latitudes, e.g., \citet{gon11a}. Importantly for our analysis, $\sigma_{RC}$ does not significantly change for the different colors and color sections analyzed, implying that there are no important correlations between colors and variations of distance and metallicity for the stellar populations that the RC trace, the main sources of systematic error in our method, mentioned in Section~\ref{intro}. Also, variations of distance and metallicity in the RC cannot certainly explain the change in more than 1 mag for the position in the peak of the gaussian fit, as a function of color, seen in Figure~\ref{fig_cmd}.

To finish our analysis, we performed a linear fit to the positions of the bulge RC as a function of color. The slope of this linear fit informs us about the selective-to-total extinction ratios (see Tables~\ref{tab_ext} and \ref{tab_extquad}). 

\section{Results}
We perform the above mentioned analysis in the different CMDs available, first keeping the $(J-K_S)$ color in one axis, and changing the magnitudes obtained in the different available filters in the other. This way changes in the slope of the selective-to-total extinction ratios are only due to variations in $A_{\lambda}$, and we are able to obtain the ratios $A_{\lambda}:A_{K_S}$ reported in Table~\ref{tab_ext}. These values are a little smaller than those provided by \citet{nis06a,nis09}, and even though our sample is more complete, engulfing every region in the galactic central area down to smallest absolute values in galactic latitude, and we cover much redder colors, we should emphasize that their photometry and calibrated extinctions are in the MKO \citep{nis06a} and 2MASS system \citep{nis09}, while ours is in the Vista system\footnote{The VISTA photometric system is tied to, but different from, the 2MASS photometric system. http://casu.ast.cam.ac.uk/surveys-projects/vista/technical/photometric-properties}. Using the mean wavelength of the VVV filters \citep{sai12b}, we observe the wavelength dependence of extinction to have a steeper distribution than previously measured for near-infrared wavelengths towards the Galactic Center (GC), following a power law $A_{\lambda}\propto{\lambda}^{-2.47}$ (see Table~\ref{tab_ext}). We should however highlight that the study by \citet{nis06a,nis09} did not include observations at wavelengths shorter than the $1.2 {\mu}m$ ($J$ filter). 

We also perform a similar analysis, but this time maintaining the $K_s$ photometry in one axis of the CMDs, and varying the colors in the other axis. As before, we can see in Table~\ref{tab_ext} that not only are our values smaller than canonical values for galactic extinction at these wavelengths \citep{car89}, as expected for these regions, but they are also smaller than those stated in previous studies of these central regions \citep{nis06a,nis09}. The measured values are more in line with those reported in \citet{alo15} from the analysis of the RRLyrae discovered in 2 highly reddened globular clusters in the inner Galaxy, also using VVV observations.

\begin{figure*}
\plotone{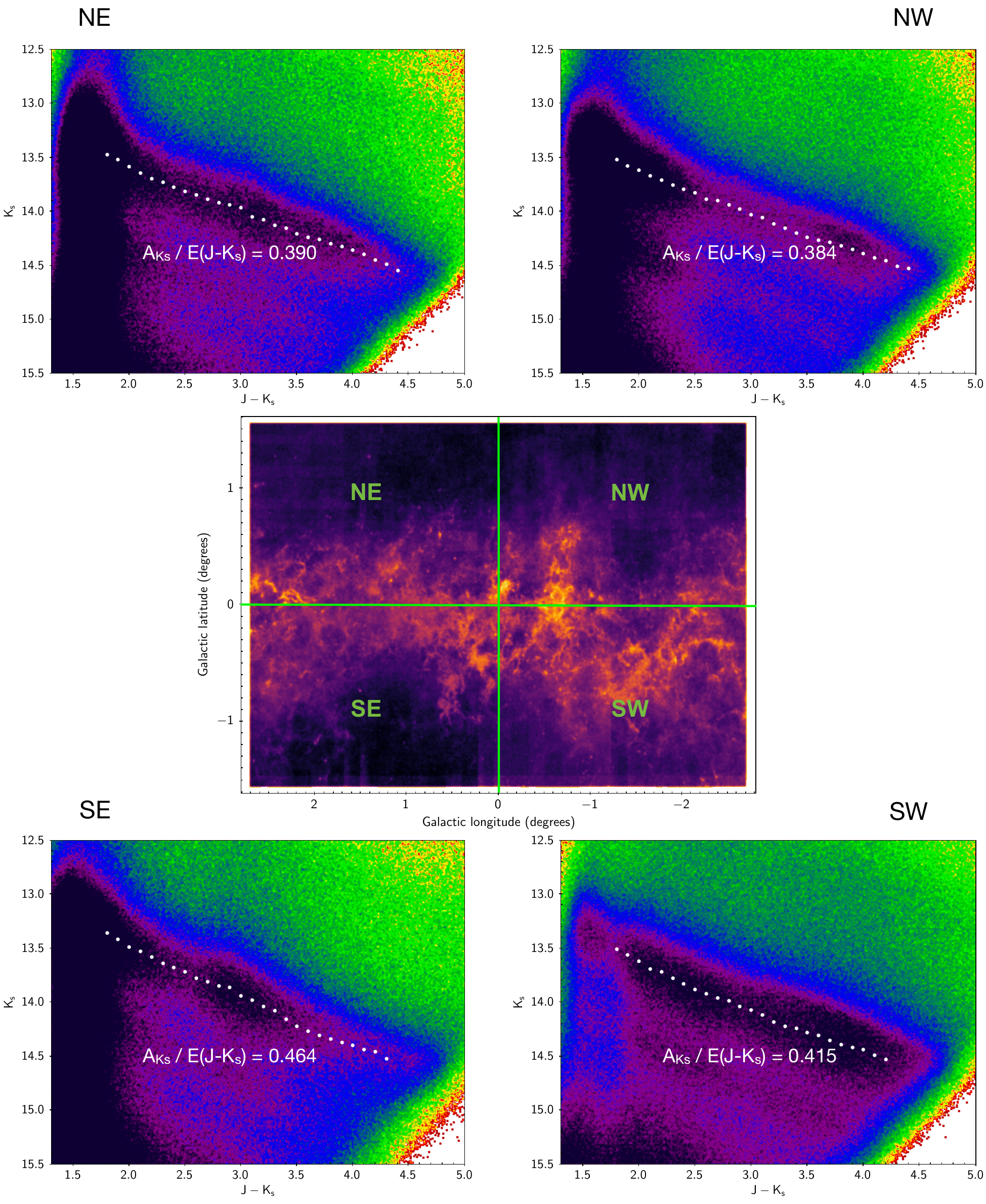}
\caption{\footnotesize In the middle, the density map of the studied region of the sky divided in four equal size quadrants. In the corners, section of the $(J-K_s)$ vs. $K_s$ CMD where most of the giant stars from the Galactic Bulge are located, for those four different quadrants. We can observe that the the reddening vector changes for every quadrant, and that northern quadrants at positive latitudes have smaller slopes in their reddening vector. As in the other figures, colder colors mean higher densities of stars. Note however that same colors mean different densities in every CMD, as the color distribution has been shifted in the CMD of every quadrant to better highlight the RC variations with differential extinction.}
\label{fig_quad}
\end{figure*}

\citet{nis06a} hints at the possibility that their reported values are not universal for the inner region, and that variations in the different quadrants in the inner region may exist. In order to investigate this possibility, we perform an additional analysis, maintaining the $K_s$ photometry in one axis and varying the colors in the other axis as before, but this time dividing the studied region in 4 equal size quadrants separated at galactic latitudes and longitudes equal to $0\degr$ (see Figure~\ref{fig_quad}). We find all the slopes to be different, and regions located at positive latitudes to have smaller slopes, and therefore, smaller selective-to-total extinction ratios than regions located at negative latitudes, as previously reported in \citet{nis09} (see Table~\ref{tab_extquad}). We include these variations in the second ${\sigma}$ term provided in Table~\ref{tab_ext}. 

\section{Conclusions}
We provide the deepest, and most complete and homogeneous atlas of a section covering a few square degrees in the central region of the Galaxy. More than 30 million sources have been resolved and precise PSF photometry in $Z$, $Y$, $J$, $H$ and $K_s$ near-infrared filters extracted using observations from the VVV survey. The CMDs containing these objects suffer from high and significantly differential reddening, especially the Galactic Bulge stars. We are able to identify the positions of RC stars in the Galactic Bulge out to very red colors ($J-K_S\sim4.5$) in the CMD, and down to absolute values of galactic latitudes $|b|=0\degr$ in the sky, unprecedented for such a big area in the inner Galaxy. The RC method allows us to study the reddening law towards these low galactic coordinates lines-of-sight. With values $A_{K_s}/E(H-K_s)$=1.104, $A_{K_s}/E(J-K_s)$=0.428, $A_{K_s}/E(Y-K_s)$=0.279, $A_{K_s}/E(Z-K_s)$=0.201, and $A_{Z}:A_{Y}:A_{J}:A_{H}:A_{K_s}$ of 7.74:5.38:3.30:1.88:1.00, we find the mean selective-to-total extinction ratios and the ratio of absolute extinctions towards the innermost Galaxy at near-infrared wavelengths to be smaller than previously believed, and extinction towards the GC to follow a distribution as a function of wavelength steeper than previously reported, consistent with a power law $A_{\lambda}\propto{\lambda}^{-2.47}$. We also found the selective-to-total extinction ratio not to be constant even in this relatively small area, but show variations that should be considered if our reported extinction ratios are used.  

\acknowledgements The authors gratefully acknowledge the use of data from the ESO Public Survey program ID 179.B-2002, taken with the VISTA telescope, and data products from the Cambridge Astronomical Survey Unit. J.A-G. acknowledge support by FONDECYT Iniciaci\'on 11150916, by the Ministry of Education through grant ANT-1656, and by the Ministry of Economy, Development, and Tourism's Millennium Science Initiative through grant IC120009, awarded to the Millennium Institute of Astrophysics (MAS). R.K.S. acknowledges support from CNPq/Brazil through projects 308968/2016-6 and 421687/2016-9.

\clearpage

\begin{deluxetable}{Cccccc}
  \tablewidth{0pc} 
  \tablecaption{Selective-to-total extinction ratios towards the GC \label{tab_ext}}
  
  \tablehead{  &
    \colhead{VVV-Red clump} &
    \colhead{Alonso-Garc\'{i}a15\tablenotemark{1}} & 
    \colhead{Nishiyama06} & 
    \colhead{Nishiyama09} & 
    \colhead{Cardelli89}} 
  
  \startdata
A_{K_s}/E(H-K_s) & 1.104$\pm$0.022$\pm$0.2 & 1.28$\pm$0.14 & 1.44$\pm$0.01 & 1.61$\pm$0.04 & 1.87 \\
A_{K_s}/E(J-K_s) & 0.428$\pm$0.005$\pm$0.04 & 0.45$\pm$0.04 & 0.494$\pm$0.006 & 0.528$\pm$0.015 & 0.72 \\
A_{K_s}/E(Y-K_s) & 0.279$\pm$0.003$\pm$0.02 & 0.23$\pm$0.02 & \nodata & \nodata & 0.43 \\
A_{K_s}/E(Z-K_s) & 0.201$\pm$0.003$\pm$0.03 & 0.15$\pm$0.02 & \nodata & \nodata & 0.31 \\
\hline
A_{H}/A_{K_s} & 1.88$\pm$0.03 & \nodata & 1.73$\pm$0.03 & 1.60$\pm$0.04 &1.54\\
A_{J}/A_{K_s} & 3.30$\pm$0.04 & \nodata & 3.02$\pm$0.04 & 2.86$\pm$0.08 &2.38\\
A_{Y}/A_{K_s} & 5.38$\pm$0.07 & \nodata & \nodata & &3.31\\
A_{Z}/A_{K_s} & 7.74$\pm$0.11 & \nodata & \nodata & &4.24\\
\hline
\alpha & 2.47$\pm$0.11 & \nodata & 1.99$\pm$0.02 & 2.00 & 1.64$\pm$0.02\\
  \enddata
  \tablenotetext{1}{Note the position of the two studied globular clusters puts them closer to the SE quadrant in Table~\ref{tab_extquad}, providing an even better agreement with the reported extinction values.}
  \tablecomments{The first reported error in the first column corresponds to the statistical error, the ${\sigma}$ of the linear fit, while the second one is the systematic error, corresponding to the variations depending on sky position reported in Table~\ref{tab_extquad}. The errors reported in the next columns correspond to statistical errors.}
\end{deluxetable}

\begin{deluxetable}{Ccccc}
  \tablewidth{0pc} 
  \tablecaption{Selective-to-total extinction ratios towards the different quadrants \label{tab_extquad}}
  
  \tablehead{  &
    \colhead{NE} &
    \colhead{SE} & 
    \colhead{NW} & 
    \colhead{SW}} 
  
  \startdata
A_{K_s}/E(H-K_s) & 1.02$\pm$0.03 & 1.30$\pm$0.03 & 0.97$\pm$0.03 & 1.21$\pm$0.05 \\
A_{K_s}/E(J-K_s) & 0.390$\pm$0.006 & 0.464$\pm$0.006 & 0.384$\pm$0.005 & 0.415$\pm$0.008 \\
A_{K_s}/E(Y-K_s) & 0.265$\pm$0.006 & 0.273$\pm$0.003 & 0.264$\pm$0.003 & 0.282$\pm$0.005 \\
A_{K_s}/E(Z-K_s) & 0.195$\pm$0.006 & 0.201$\pm$0.003 & 0.178$\pm$0.004 & 0.228$\pm$0.006  \\
  \enddata
\end{deluxetable}


\end{document}